\documentclass[12pt]{article}
\usepackage[dvips]{graphicx}
\usepackage{subfigure}

\usepackage{epsfig,amsmath,amssymb,verbatim,mathrsfs}

\def\beq{\begin{eqnarray}}
\def\eeq{\end{eqnarray}}
\def\bea{\begin{eqnarray}}
\def\eea{\end{eqnarray}}

\newcommand{\be}{\begin{equation}}
\newcommand{\ee}{\end{equation}}

\begin{document}

\setlength{\baselineskip}{0.2in}


\begin{titlepage}
\noindent
\vspace{0.2cm}

\begin{center}
  \begin{Large}
    \begin{bf}
Comments on the Hierarchy Problem in \\ Effective  Theories
     \end{bf}
  \end{Large}
\end{center}

\begin{center}

\begin{large}

{ Archil~Kobakhidze\footnote{archilk@physics.usyd.edu.au} and Kristian~L.~McDonald\footnote{klmcd@physics.usyd.edu.au}} \\
     \end{large}
\vspace{0.5cm}
  \begin{it}
ARC Centre of Excellence for Particle Physics at the Terascale\\
\vspace{0.3cm}
School of Physics, The University of Sydney, NSW 2006, Australia\\\vspace{0.5cm}

\end{it}

\end{center}


\begin{abstract}
We discuss aspects of the hierarchy problem in effective theories with light scalars and a large, physical ultraviolet (UV) cutoff. We make two main points: (1) The \emph{(naive) fine-tuning} observed in an effective theory does not automatically imply that the UV completion is fine tuned. Instead, it gives a type of upper bound on the severity of the \emph{actual tuning} in the UV completion; the actual tuning can be less severe than the naive tuning or even non-existent. (2) Within an effective theory, there appear to be two types of parameter relations that can alleviate the sensitivity of the scalar mass to the cutoff --- relationships among dimensionless couplings or relationships among dimensionful parameters. Supersymmetric models provide symmetry-motivated examples of the former, while scale-invariant models give symmetry-motivated examples of the latter.

\end{abstract}

\vspace{1cm}

\end{titlepage}

\setcounter{footnote}{0}
\setcounter{page}{1}


\vfill\eject


\section{Introduction}
The hierarchy problem~\cite{Wilson:1970ag} has driven much research into physics beyond the Standard Model (SM) in recent decades. When taken as an effective theory with a large cutoff, it is difficult to understand how the Higgs boson remains light relative to the large cutoff effects. With the experimental discovery of the Higgs boson at the LHC, the desire to understand how scalars can remain light, despite large cutoff-dependent corrections, has gained urgency in recent years~\cite{Dubovsky:2013ira}-\cite{Foot:2013hna}.

In this work we discuss aspects of the hierarchy problem in effective theories with light scalars and  a large physical ultraviolet (UV) cutoff. We discuss two main points. Firstly, we note that the \emph{(naive) fine-tuning} found in an effective theory does not automatically imply that the UV completion is fine tuned. Rather, it gives a sort of upper bound on the severity of the \emph{actual tuning} present in the underlying UV completion; the actual tuning due to the new physics at the cutoff scale can be less severe than the naive tuning or even non-existent. Secondly, we note that within an effective theory there appear to be two types of parameter relations that can alleviate the sensitivity of the scalar mass to the cutoff; a relationship among dimensionless couplings \emph{\'a~la} the Veltman condition, or  a relationship among dimensionful parameters. Supersymmetric (SUSY) models give symmetry-motivated examples in which a technically-natural Veltman-like condition arises, while scale-invariant models are symmetry-motivated examples where relations between dimensionful parameters are expected.

In essence, these two points pertain to the two different aspects of the hierarchy problem, namely the naturalness of new particle thresholds and the tuning associated with ``pure cutoff" effects. If the UV physics takes the form of new thresholds, or behaves similar to such, the naive tuning can overestimate the actual tuning. This has implications for model-building approaches to beyond-SM physics~\cite{Farina:2013mla,Foot:2013hna}. However, if the cutoff accurately represents the behaviour of the UV physics, one expects a natural theory to shield the infrared (IR) sector. These points are elaborated within.

The layout of this paper is as follows. In Section~\ref{sec:tuning} we discuss the naive tuning in an effective theory and consider a simple example where the actual tuning in the UV completion can be less severe. In Section~\ref{sec:hierarchy} we consider simple relationships among parameters in an effective theory that can shield the IR sector from cutoff effects. Conclusions are drawn in Section~\ref{sec:conc}.
\section{Naive-Tuning Versus Actual-Tuning\label{sec:tuning}}
Consider the effective theory for a self-interacting scalar field $S$, with Lagrangian
\bea
\mathcal{L}_S^\Lambda&=& \partial^\mu S^*\partial_\mu S-\bar m^2_S(\Lambda)|S|^2 - \lambda_S |S|^4 +\ldots \label{eq:S_eff_lagrange}
\eea
The parameters depend on the cutoff scale $\Lambda$, and the dots denote  non-renormalizable irrelevant IR operators with mass-dimension $d>4$, suppressed by factors of $\Lambda^{4-d}$. Calculating  the one-loop corrected mass in the effective theory gives\footnote{We suppress numerical loop-factors in this section.}
\bea
m_{S}^2&=&\bar m^2_S(\Lambda)+\delta m_\Lambda^2\ \equiv\ \bar m^2_S(\Lambda) +  \lambda_S \{\Lambda^2 + \bar m^2_S(\Lambda) \log(\bar m^2_S(\Lambda)/\Lambda^2)\}\ .\label{light_mass_Ir}
\eea
For large values of $\Lambda$, the theory appears to have a hierarchy problem, with small values of $m_S^2\ll \Lambda^2$ requiring a fine-tuning between $\bar m_S(\Lambda)$ and $\Lambda$. The effective theory is said to be fine-tuned because the origin of this cancellation is understood within the effective theory. 

It is important to distinguish two qualitatively different contexts in which Eq.~(\ref{light_mass_Ir}) is interpreted. In the first case, $\Lambda$ is merely as a tool to regularize the divergent loop-integral. Then, both $\Lambda$ and the  $\Lambda$-dependent bare parameters are regarded as unphysical quantities. In fact, one must take the limit $\Lambda\to\infty$ and remove all divergences by renormalizing the unphysical bare masses and couplings, leaving finite physical parameters. From the symmetry perspective, a theory with an unphysical cut-off possesses a scale invariance that is \emph{softly} broken by the explicit mass terms and the logarithmic quantum corrections~\cite{Bardeen:1995kv}. 
The softness of the breaking is reflected in the infrared fixed-point structure of the mass RGEs~\cite{Wetterich:1983bi}.  

The second interpretation of Eq.~(\ref{light_mass_Ir}), of interest here, applies when the cut-off $\Lambda$ is physical and therefore associated with a new 
scale in the UV-completion of the effective theory. In this case the effective theory has a hierarchy problem~\cite{Wilson:1970ag}, and the cut-off $\Lambda$, as well as the $\Lambda$-dependent bare parameters and the set of irrelevant operators, fully encode information about the UV theory. Absent knowledge of the underlying theory, one cannot renormalize away the $\Lambda$-dependence.  We refer to the fine-tuning associated with this hierarchy problem as the ``naive fine-tuning of the effective theory," or more simply as ``the naive tuning." For a given fixed value of $\lambda_S=\mathcal{O}(1)$ (for example), this naive tuning requires two parameters of $\mathcal{O}(\Lambda^2)$ to cancel out at a precision of $\mathcal{O}(m_S^2/\Lambda^2)$. With $m_S^2\ll \Lambda^2$ this tuning is severe. 

One can phrase the hierarchy problem of Eq.~(\ref{light_mass_Ir}) in the following way. Consider  values of the parameters in the effective theory that generate some fixed value for the  scalar mass ${m}_S^2$. Now shift the bare mass $\bar m^2_S(\Lambda)$ as follows:
\bea
\bar m^2_S(\Lambda)\rightarrow \bar m^2_S(\Lambda)+\delta\bar m^2_S(\Lambda), \quad\mathrm{with}\quad \frac{\delta\bar m^2_S(\Lambda)}{\bar m^2_S(\Lambda)}\lesssim\mathcal{O}(1).\label{eq:bare_mass_change1}
\eea
This will, in general, induce a shift in the scalar mass:
\bea
{m}_S^2\rightarrow {m}_S^2 +\delta{m}_S^2.
\eea
The effective-theory has a  hierarchy problem if $\delta{m}_S^2/{m}_S^2\gg \mathcal{O}(1)$ ---~i.e.~if small changes in the bare mass create a large change in the scalar mass. For generic couplings, an effective theory with $m_S^2\ll\Lambda^2$ is expected to have a hierarchy problem. One must fine-tune $\bar m^2_S(\Lambda)\simeq\mathcal{O}(\Lambda^2)$ against the $\Lambda^2$-term to enable $m_S^2\ll\Lambda^2$. Thus, shifts of $\delta\bar m^2_S(\Lambda)/\bar m^2_S(\Lambda)=\mathcal{O}(1)$ give  $\delta\bar m^2_S(\Lambda)=\mathcal{O}(\Lambda^2)$, which in turn gives
\bea
{m}_S^2\rightarrow {m}_S^2 + \mathcal{O}(\Lambda^2)\gg m_S^2.
\eea
The hierarchy problem manifests through this extreme sensitivity of the physical scalar mass to  small changes in the  effective-theory mass-parameter.

With knowledge of the UV completion for the theory $\mathcal{L}_S^\Lambda$, one can investigate the origin of the naive tuning by calculating the ``actual tuning" in the UV completion. One would like to know if the naive tuning accurately encodes the actual tuning. Our main point in this section is that the actual tuning can be less severe than the naive tuning and in some cases may even be absent. One should think of the naive tuning as representing a worst case scenario for the severity of the tuning associated with the new physics at the scale $\Lambda$. We demonstrate this with a simple example.

The scale $\Lambda$ is assumed physical, in the sense that new physics appears at this scale. There are, in principle, two classes of UV completions that one could consider. In one class, the scalar $S$ persists as a physical degree of freedom beyond the scale $\Lambda$, and the new physics takes the form of additional degrees of freedom with mass $M_H\sim\Lambda$. A typical example occurs when the theory is UV completed by a new heavy scalar, $H$. If the two scalars couple via a quartic interaction, $\lambda_{mix}S^2H^2$, the UV completion generates a mass correction for the light scalar of the form
\bea
\delta m^2_{S,H}&=& \lambda_{mix} M_H^2 \log (M_H^2/\mu^2),
\eea
where $\mu$ denotes a renormalization scale in the UV completion.\footnote{For our purpose in this section it suffices to take the cutoff for the UV theory merely as a regulator. We further  discuss cutoff effects  in the next section.} With regard to fine-tuning due to $H$, one can differentiate three cases:
\begin{itemize}
\item For $\lambda_{mix}\sim\mathcal{O}(1)$, there is a hierarchy problem and a light scalar with $m_S\ll M_H$ requires a fine-tuned UV completion. The required tuning  is at the level of $\mathcal{O}(m_S^2/\Lambda^2)$, occurring between  quantities of $\mathcal{O}(\Lambda^2)$. In this case the actual tuning agrees with the naive tuning of the IR observer.

\item For  $m_S^2/M_H^2\ll\lambda_{mix}\ll1$, there is still a hierarchy problem and the UV completion remains fine-tuned. However, now the actual tuning  is at the level of $\mathcal{O}(m_S^2/[\lambda_{mix}M_H^2])$, between quantities of $\mathcal{O}(\lambda_{mix}M_H^2)\ll\Lambda^2$. The actual tuning due to the physics at $\Lambda$ is less severe than the naive tuning.

\item For $\lambda_{mix}\lesssim m_S^2/M_H^2$, there is no hierarchy problem. The loop-correction does not exceed the physical scalar mass and the theory is technically natural. This case is contrary to the naive expectation of the IR observer; the naively-tuned effective-theory possesses a technically-natural UV completion.
\end{itemize}

Note that the second and third cases are only possible if values of $\lambda_{mix}\ll1$ are technically natural. This is true if, e.g., the limit $\lambda_{mix}\rightarrow0$ gives an enhanced Poincar\'e symmetry~\cite{Foot:2013hna}, as occurs here. This example shows that the naive tuning in an effective theory can be more severe than the actual tuning due to the physics at the scale $\Lambda$. In Appendix~\ref{app:SM} we briefly discuss the above cases in the context of the SM.

In the other class of UV completions for $\mathcal{L}_S^\Lambda$, there is a transition to a new theory at the scale $\Lambda$, such that the degrees of freedom are different and the scalar $S$ ceases to exist. For example, this occurs if $S$ is a low-energy composite object\footnote{In general, one would expect more IR composite states than a single scalar for this case.}  and the transition to more-fundamental ``quarks" occurs at $\Lambda$, or similarly if $\Lambda$ is a minimal length-scale in the UV theory,\footnote{Analogous to the role played by the inter-atomic spacing when describing spin-correlation functions of a magnetic system by an effective scalar field theory.} or a type of non-locality scale such as in string theory. In these examples the actual tuning required in the UV completion is expected to be as severe as the naive tuning.

Note that an IR observer cannot differentiate between these two classes of UV completions, absent information about the UV physics. When faced with a naively-tuned effective theory an IR observer can, at best, conclude that the underlying UV completion \emph{may} be fine tuned, due to new physics at the scale $\Lambda$. The naive tuning provides a type of upper bound for the severity of the actual tuning  in the UV completion as a result of the new physics at the scale $\Lambda$; the actual tuning in the completion can be less severe or in some cases even non-existent. 

The physical origin of the fine-tuning in the two classes of UV completions is distinct. The $\Lambda^2$-term in \eqref{light_mass_Ir} encodes the mass correction to the IR scalar $S$, due to the UV (Euclidean) momentum modes with  $|p_{\rm E}|\sim\Lambda$ for the IR degree of freedom. In the first class of models, where heavy new physics  with mass $M_H\sim\Lambda$ appears, the origin of the actual tuning differs from the source of the naive tuning. The actual tuning, if present, is due to mass corrections from the heavy UV physics, while the naive tuning results from UV momentum modes for the IR fields. These two effects have distinct physical meanings and this is why the naive tuning and the actual tuning can differ. One should think of the naive tuning as being a proxy for the actual tuning --- the existence of a naive tuning in the IR theory indicates that the UV completion may contain an actual tuning. Note that, from the perspective of the UV theory, there is nothing special about modes with $|p_{\rm E}|\sim\Lambda$ for the IR scalar; these just happen to have the same momentum as the new physics scale. 

In the case where $S$ does not persist in the UV, the momentum modes with $|p_{\rm E}|\sim\Lambda$ for the IR scalar $S$ are ``special" in the sense that modes with $|p_{\rm E}|>\Lambda$ simply do not exist.  Now the $\Lambda^2$-term has a clear physical meaning and one understands why the modes $|p_{\rm E}|\sim\Lambda$ would give a mass correction of greater physical significance than modes with,~e.g.,~ $|p_{\rm E}|\sim \Lambda'\ll \Lambda$. In this case there is no obvious reason why the  bare mass and the cutoff should cancel-out so precisely; the natural value for the scalar mass is expected to be $m_S\sim\Lambda$. Said differently, the naive tuning is expected to well-approximate the actual-tuning.

\section{Removing the Cutoff via Parameter Relations\label{sec:hierarchy}}
We now turn to a different aspect of the hierarchy problem. Having focused mainly on new threshold effects in the preceding, we now focus on the $\Lambda^2$-term, assuming that it encodes a real physical effect that must be dealt with, for the theory to be natural. We consider the simplest possibilities for alleviating the cutoff sensitivity within the effective theory, namely by parameter relations that shield the  light scalar mass from cutoff effects. To discuss this matter it is helpful to consider a more detailed IR sector. 

Consider the  effective theory for a system of  real scalars $S=(s_1,s_2,...)^{\rm T}$, 
gauge fields $V$, and Weyl fermions $F$,  assumed  valid up to a large UV cut-off $\Lambda$, which is understood in the Wilsonian sense. If present, a hierarchy problem would manifest in the relevant operators $S_aS_b$ that appear in the quantum-corrected effective action. 
The coefficients of these operators are given by the non-derivative part of the 2-point functions 
$\Gamma^{(2)}_{ab}$. In the 1-loop approximation they are:
\begin{eqnarray}
  \left({m}_S^2(\Lambda, \mu)\right)_{ab}
= \left(\bar m_S^2(\Lambda)\right)_{ab}+\sum_{A=S,V,F}(-1)^{2J_A}\left(2J_A+1\right)\frac{\left(g_A\right)_{abcd}}{16\pi^2}
\left[\Lambda^2\delta_{cd}-\left(\bar m_A^2(\Lambda)\right)_{cd} 
\ln\frac{\Lambda^2}{ \mu^2}
\right],\nonumber\\
\label{s2-1}
\end{eqnarray} 
where $\bar m^2_A(\Lambda)$ is the effective   bare-mass for the field $A$ of spin $J_A$, and $\mu$ is an arbitrary renormalization scale, $|m_A|<\mu <\Lambda$ . Here $g_A$ denotes the matrix of dimensionless couplings between the field  $A$ and the scalars $S$, defined through  the interaction terms as
\bea
\frac{1}{4!}(g_S)_{abcd}S_aS_bS_cS_d, \quad \frac{1}{2}(g_V)_{abcd}V_{a}V_{b}S_{c}S_{d} \quad\mathrm{and}\quad (y_F)_{abc}\bar F_aF_bS_c,
\eea
where $(y_{F})_{abk}(y_{F})_{cdk}=(g_{F})_{abcd}$.  As before, for light values of ${m}_S^2\ll\Lambda^2$, the effective theory has a hierarchy problem.

How can one remove this sensitivity to the cutoff? From the perspective of the effective theory, there appear to be two types of relationships that could remove the cutoff sensitivity in Eq.~\eqref{s2-1}. One could consider a relationship among dimensionless couplings that cancels the $\Lambda^2$-term, or a relationship among dimensionful quantities that relates the bare mass to the UV scale. For such a relationship to provide a viable explanation (i.e.~not transfer the tuning to a different sector), it should be motivated by a symmetry. We discuss these two cases in turn.

\subsection{Relationships Among Dimensionless Couplings\label{sec:dim_less_coupling}}
First  consider the case where the $\Lambda^2$-dependence is removed by a relationship among dimensionless couplings. We discuss two examples, differentiated by the absence/presence of an underlying symmetry.
\paragraph{The Veltman Condition.}
The $\Lambda^2$-term in Eq.~\eqref{s2-1} disappears if the dimensionless couplings in the theory are related:
\begin{equation}
\sum_{A=S,V,F}(-1)^{2J_A}\left(2J_A+1\right)(g_A)_{abcc}=0.
\label{eq:veltman_relation}
\end{equation}
This possibility was introduced by Veltman~\cite{Veltman:1980mj}, and studied prior-to (after) the Higgs discovery in Ref.~\cite{Jack:1989tv}  (\cite{Hamada:2012bp}). In the SM, this can be converted into a mass relation:
\bea
m_h^2 +2m_W^2+m_Z^2-4m_t^2=0,
\eea
giving $m_h\sim300$~GeV, in conflict with the data. However, this relation might be satisfied in the UV, perhaps with additional beyond-SM fields also participating~\cite{Hamada:2012bp}. If the Veltman condition is realized, the $\Lambda^2$-term in Eq.~\eqref{s2-1} cancels out and the bare Higgs mass is similar to the observed mass, $m_h^2/\bar{m}_h^2(\Lambda)=\mathcal{O}(1)$. Consequently,  small  changes to the bare mass  give small changes to the Higgs mass, and the theory appears natural. However, the (generalized) Veltman condition is not motivated by any symmetry. From the perspective of the effective theory, one cannot understand why such a relationship exists, nor how it remains radiatively stable. If a Veltman-like relation holds, the hierarchy problem is revealed by looking at small shifts in individual dimensionless couplings. Under  a small shift $g_A\rightarrow g_A+\delta g_A$ for a particular coupling $g_A$,  the cancellation of the $\Lambda^2$-term ceases to function and the need for fine-tuning is manifest. Thus, the hierarchy problem of the effective theory is not resolved; one has simply exchanged a tuning among mass-parameters for a tuning  amongst couplings.


\paragraph{Supersymmetry.} Models with exact SUSY possess an equal number of bosonic and fermionic degrees of freedom (i.e., $\sum_{A=S,V,F}(-1)^{2J_A}\left(2J_A+1\right)=0$). Furthermore, SUSY forces relationships among dimensionless coupling constants, and requires multiplets to be mass-degenerate, namely   
\begin{equation}
\sum_{A=S,V,F}(-1)^{2J_A}\left(2J_A+1\right)(g_A)_{abcc}=0
\label{s2-7}
\end{equation}
and 
\begin{equation}
\sum_{A=S,V,F}(-1)^{2J_A}\left(2J_A+1\right)(g_A)_{abcd}(m_A)_{cd}^2=0.
\label{s2-8}
\end{equation}
These equations  reflect the perturbative non-renormalization theorem~\cite{Grisaru:1979wc}, according to which  only  wavefunction renormalization is required in N=1 SUSY theories. Eq.~(\ref{s2-7})  solves the hierarchy problem because it forces the coefficient of the $\Lambda^2$-term to vanish in Eq.~(\ref{s2-1}). Note the similarity with the Veltman condition; it is clear that SUSY models can be thought of as symmetry-motivated examples where a Veltman-like condition is automatically achieved.

In realistic applications,  effective theories describing supersymmetric extensions of the SM cannot possess exact SUSY. However, Eq.~(\ref{s2-7}) also holds  in \emph{softly}-broken SUSY theories, since  the  
soft SUSY-breaking  terms are dimensionful parameters that do not appear in (\ref{s2-7}). 
On the other hand,  Eq.~\eqref{s2-8} is modified because the soft-breaking terms  lift the fermion-boson mass degeneracy, 
and set the SUSY-breaking scale, 
\bea
\sum_{A=S,V,F}(-1)^{2J_A}\left(2J_A+1\right)(g_A)_{abcd}(m_A)_{cd}^2\sim M^2_{SUSY}.
\eea 
Then, provided $M_{SUSY}\ll\Lambda$, light scalars with mass $m_S^2=\mathcal{O}(M_{SUSY}^2)$ are technically natural in softly-broken SUSY models. In terms of the bare scalar mass, this corresponds to $\bar m^2_S(\Lambda)\lesssim\mathcal{O}(M_{SUSY})$, so a small shift  induces the change $\delta m^2_S\lesssim\mathcal{O}(M_{SUSY}^2)$ in the physical mass, manifesting naturalness. 

If  SUSY is broken in the UV completion, then presumably the  fermion-boson mass degeneracy is broken by large $\mathcal{O}(\Lambda)$ effects in some heavy (hidden) sector of the theory. The light (visible) sector must couple to the heavy sector with sufficiently weak couplings to enable $m_S^2\sim M^2_{SUSY}\ll \Lambda^2$. This is  what happens in most of the realistic particle physics models, where SUSY is spontaneously broken at a high energy scale in a heavy hidden sector,  and feebly communicated to the visible sector.     

In the case of \emph{hard} SUSY-breaking, Eq.~(\ref{s2-7}) is also violated and the SUSY-breaking scale is set by
\bea
\sum_{A=S,V,F}(-1)^{2J_A}\left(2J_A+1\right)(g_A)_{abcc}\Lambda^2 \sim M^2_{SUSY}.
\eea 
Light scalars can only be accommodated  if the \emph{hard} SUSY-breaking dimensionless couplings are sufficiently small, i.e. the relation (\ref{s2-7}) is satisfied with sufficient accuracy. We  stress  that \emph{hard} SUSY-breaking also allows technically-natural light scalars, as the limit when Eq. (\ref{s2-7}) is strictly satisfied corresponds to an increased symmetry in the theory. In fact, small, \emph{hard} SUSY-breaking terms frequently appear in SUSY models without causing problems.

Recall that, although the $\Lambda^2$-term canceled out in models with a Veltman condition, the hierarchy problem manifested under small shifts to the dimensionless couplings, $g_A\rightarrow g_A+\delta g_A$. SUSY cures this problem by demanding that the shift $g_A\rightarrow g_A+\delta g_A$ is accompanied  by complimentary shifts $g_A'\rightarrow g_A'+\delta g_A'$ in any couplings related to $g_A$ through SUSY. One must enforce a relationship among dimensionless couplings to ensure that Eq.~\eqref{s2-7} is satisfied. To vary a single coupling without varying the SUSY-related couplings would amount to a departure from the physical content of the theory. This symmetry-motivated origin for the coupling-relation ensures its stability under radiative corrections and tells us the couplings are likely born in some related way.

To summarize, in the presence of two (or more) sectors with hierarchically different masses, SUSY can ensure stability of the hierarchy if it is broken softly with all the soft-breaking mass parameters being of order $M_{SUSY}\ll\Lambda$, as in softly-broken SUSY GUTs. In other cases, the heavy hidden sector should couple to the light visible sector very weakly. The latter case may or may not be technically natural, depending on details of the model~\cite{Foot:2013hna}.

\subsection{Relationships Among Dimensionful Couplings\label{sec:SI_hierarchy}}
SUSY models and the Veltman-condition both cancel out the quadratic divergences via a relationship among dimensionless couplings. The other possibility is that the sensitivity to the cutoff in Eq.~\eqref{s2-1} is alleviated by a relationship among dimensionful parameters in the effective theory. Indeed, the hierarchy problem of the SM  manifests as a tuning between the bare scalar-mass and the UV scale --- this sensitivity between the \emph{only two} dimensionful parameters of the effective theory may suggest a deeper connection.

\paragraph{Relationship without symmetry.} In analogy with the Veltman condition, one can imagine a UV completion that triggers a relationship between the dimensionful parameters in the effective theory. However, for arbitrary relations among dimensionful parameters one cannot understand the origin of the relation within the effective theory, and the relation is equivalent to a tuning. For example, if the UV completion triggers the relation  
\begin{equation}  
({\bar m}_S^2(\Lambda))_{ab}+\sum_{A=S,V,F}(-1)^{2J_A}\left(2J_A+1\right)\frac{1}{16\pi^2}(g_A)_{abcc} \Lambda^2=M_{ab}^2,
\label{eq:dimful_veltman}
\end{equation} 
for some fixed $M_{ab}^2\ll \Lambda^2$, this would ``remove" the quadratic divergence. However, the IR observer could not distinguish this from a fine-tuning --- this relation is  the standard expression for the fine-tuning! As with the Veltman condition, one cannot understand how such a relationship remains stable under radiative corrections in the effective theory framework, so the tuning persists. This is evidenced by the fact that a small shift in the bare-mass induces a large shift in the physical mass.

\paragraph{Scale invariance.} Within an effective theory, scale-invariance appears to be badly broken. In addition to the bare masses $\bar m_A(\Lambda)$, and the logarithmic quantum-anomalous terms, which break scale-invariance softly, one  encounters \emph{hard}-breaking relevant operators $\sim \Lambda^2$, which introduce the  quadratic sensitivity of the light masses to the cut-off scale.  Nevertheless, the effective theory can describe an underlying UV theory that maintains scale invariance, at least at the classical level. 
The renormalized mass terms computed (within the perturbative framework) 
in a scale-invariant theory are necessarily  zero because all bare mass terms are absent. 
On the other hand, the mass terms computed in a scale-invariant UV theory should match the corresponding mass terms $m_S^2(\Lambda,\mu)$, computed 
in the low-energy effective theory, at the matching scale 
  defined by the effective theory cut-off, $\mu=\Lambda$~\cite{Kobakhidze:2013uoa}.\footnote{A number of works studied scale-invariance  in relation to the hierarchy problem~\cite{Meissner:2006zh}-\cite{Shaposhnikov:2008xi}, and discussion of quadratic divergences appears in Ref.~\cite{Meissner:2007xv}.}  The IR effective theory for such UV theories automatically satisfies the relation,
\begin{equation}  
({\bar m}_S^2(\Lambda))_{ab}+\sum_{A=S,V,F}(-1)^{2J_A}\left(2J_A+1\right)\frac{1}{16\pi^2}(g_A)_{abcc} \Lambda^2=0,
\label{s2-9}
\end{equation}   
 ensuring the effective theory accurately matches the UV theory at the matching scale $\Lambda$. Thus, only a logarithmic sensitivity to the UV scale remains:
 \begin{equation}
(m_S^2(\Lambda, \mu))_{ab} = \sum_{A=S,V,F}(-1)^{2J_A}\left(2J_A+1\right)\frac{1}{16\pi^2}( g_A)_{abcd}(m_A^2)_{cd} \ln 
\frac{ \Lambda^2}{\mu^2}\ .
\label{s2-10} 
 \end{equation}

Eq.~\eqref{s2-9} is a concrete example of a symmetry-motivated relationship among dimensionful parameters that removes the quadratic cutoff dependence of the scalar mass. In the previous section, the hierarchy problem in the effective theory was revealed by varying the bare mass while keeping the UV scale and the dimensionless couplings fixed; small changes in the bare mass induced large changes in the physical mass. However, in a scale-invariant theory, a shift made to the bare mass while keeping the other parameters fixed amounts to a departure from the physical content of the UV theory. That is, if the UV and IR scales are born in some common way, due to an underlying scale invariance, the shift
\bea
\bar m^2_S(\Lambda)\rightarrow \bar m^2_S(\Lambda)+\delta\bar m^2_S(\Lambda), \quad\mathrm{with}\quad \frac{\delta\bar m^2_S(\Lambda)}{\bar m^2_S(\Lambda)}\lesssim\mathcal{O}(1),\label{eq:bare_mass_change_SI}
\eea
also requires a shift
\bea
\Lambda^2\rightarrow \Lambda^2+\delta\Lambda^2, \quad\mathrm{with}\quad \frac{\delta\Lambda^2}{\Lambda^2}\lesssim\mathcal{O}(1),\label{eq:bare_mass_change}
\eea
to ensure an accurate low-energy effective-theory description. If the couplings are held fixed, a small change in the bare mass of the form \eqref{eq:bare_mass_change_SI} must come partnered with a small change in the UV scale of the form \eqref{eq:bare_mass_change} to ensure that Eq.~\eqref{s2-9} holds. More generally, the shift in the bare mass should be partnered with a compensating shift in the couplings and/or the UV-scale to ensure they remain related by Eq.~\eqref{s2-9}. Only then does the effective theory accurately encode the relationship between the two scales that is inherent in the parent theory.

Note that  the soft-breaking masses $\bar m_A$ in scale invariant theories emerge from the \emph{spontaneous breaking} of scale invariance, the mechanism known as dimensional transmutation~\cite{Coleman:1973jx}. Thus, for a single source of symmetry breaking, all such masses are proportional to the scale  of this breaking, $M_{SI}$. As with SUSY models, we again encounter two distinct possibilities: (i) $M_{SI}\ll \Lambda$ and all the dimensionless couplings, and hence masses, are of the same order of magnitude~\cite{Foot:2007as}, or (ii) $M_{SI}\sim \Lambda$ and a hierarchy of masses exists due to a hierarchy among dimensionless coupling-constants; that is, hierarchically separated sectors of the theory interact sufficiently weakly to preserve the hierarchical scales~\cite{Foot:2013hna,Foot:2007iy}.\footnote{Recent works have focused on threshold effects, assuming the quadratic divergences are dealt with by an as-yet unknown mechanism~\cite{Farina:2013mla,Foot:2013hna} (the so-called ``miraculous cancellation"~\cite{Giudice:2013yca}). Scale-invariance provides a symmetry-based rationale for the neglect of quadratic divergences, offering a motivation for this perspective.}

Finally, we note the similarity between the SUSY  and scale-invariant narratives. If the SM is  an effective theory that is UV completed by the MSSM, new symmetry-motivated degrees of freedom appear at the cutoff scale to enable a Veltman-like condition among dimensionless couplings. This protects the weak scale from quadratic divergences. In the scale-invariant case, a new scale appears in the UV which enables a symmetry-motivated relationship among dimensionful parameters, ensuring the IR scale of the effective theory is related to the UV scale. Quantization of course breaks scale-invariance softly via $\ln \Lambda$ operators, but this only affects marginal, $d=4$ operators in the low-energy effective theory. 



\section{Conclusion\label{sec:conc}}
In this work we discussed aspects of the hierarchy problem in effective theories with a light scalar. We sought to make two main points, namely: (1) The naive tuning in an effective theory can be less severe than the actual tuning in the UV completion; the naive tuning gives a type of upper bound for the severity of the actual tuning associated with the new physics at the cutoff scale. (2) There appear to be two classes of parameter relations that can alleviate the quadratic cutoff-dependence in an effective theory; relations among dimensionless couplings (of the Veltman type), which ensure that the quadratic divergences cancel out, or relations among dimensionful parameters, which indicate the IR and UV scales are related through some common birth. SUSY can be thought of as a symmetry-motivated example that generates a natural relation of the former type, while  scale-invariant UV completions provide symmetry-motivated examples of the latter type. 

\section*{Acknowledgements\label{sec:ackn}}
The authors thank R.~Foot and R.~Volkas. This work was supported in part by the Australian Research Council.
\appendix
\section{Comments on the Standard Model\label{app:SM}}
We can relate the observations of Section~\ref{sec:tuning} to the hierarchy problem in the SM (treated as an effective-theory, valid up to a large UV-scale $\Lambda$). The three cases correspond to the following classes  of UV completions for the SM:
\begin{itemize}
\item $\lambda_{mix}\sim\mathcal{O}(1)\Longrightarrow$ The UV completion contains  new physics at the scale $\Lambda$, with  $\mathcal{O}(1)$ couplings to the SM. This case has a hierarchy problem and requires fine-tuning. The actual  tuning is as severe as the naive tuning in the effective theory.  The prototypical example is the UV completion of the SM by a renormalizable Grand Unified Theory, with unification scale $M_{GUT}\sim\Lambda\gg 10^2$~GeV.

\item $m_h^2/\Lambda^2\ll\lambda_{mix}\ll1\Longrightarrow$ The UV completion contains heavy physics that is weakly coupled to the SM, yet gives a radiative correction to the Higgs mass that exceeds the observed value. This case has a hierarchy problem and requires an unnatural fine-tuning. However, the actual tuning of the parent theory is less severe than the naive tuning, being between parameters of $\mathcal{O}(\lambda_{mix}\Lambda^2)\ll\Lambda^2$, rather than parameters of $\mathcal{O}(\Lambda^2)$. The prototypical example is the UV completion of the SM by heavy right-handed neutrinos, to generate neutrino masses via the seesaw mechanism,  with heavy Majorana masses in the range\footnote{The upper bound results from the standard seesaw expression, assuming $\mathcal{O}(1)$ Dirac Yukawa-couplings and that the SM neutrinos have masses $m_\nu\sim0.1$~eV.} $10^{12}\gg (M_R/\mathrm{GeV})\gg 10^7$.  Radiative corrections to the Higgs mass then exceed the observed value, necessitating a tuning between parameters of $\mathcal{O}(y_\nu^2 M_R^2)\ll\Lambda^2$, where $y_\nu^2\Leftrightarrow\lambda_{mix}$ is the Dirac Yukawa-coupling.

\item $\lambda_{mix}\lesssim m_h^2/\Lambda^2\Longrightarrow$ The UV completion contains weakly-coupled UV physics that does not generate a hierarchy problem. This case is technically-natural, provided the limit $\lambda_{mix}\rightarrow0$ decouples the UV physics from the SM~\cite{Foot:2013hna}. There is no actual-tuning in the parent theory despite the naive-tuning found in the IR. The prototypical example is the UV completion of the SM by heavy right-handed neutrinos, giving neutrino mass via the seesaw mechanism,  with heavy Majorana masses $ (M_R/\mathrm{GeV})\lesssim 10^7$~\cite{Foot:2013hna,Vissani:1997ys}. Radiative corrections to the Higgs mass are less than the observed value, and the theory is devoid of tuning. Another example is the invisible axion model, which can solve the strong CP-problem and give a dark matter candidate~\cite{Foot:2013hna,Volkas:1988cm,Farina:2013mla}.

\end{itemize}



\begin{thebibliography}{99}

\bibitem{Wilson:1970ag} 
  K.~G.~Wilson,
  Phys.\ Rev.\ D {\bf 3}, 1818 (1971);
  L.~Susskind,
  Phys.\ Rev.\ D {\bf 20}, 2619 (1979);
  S.~Weinberg,
  Physica A {\bf 96}, 327 (1979).


\bibitem{Dubovsky:2013ira} 
  S.~Dubovsky, V.~Gorbenko and M.~Mirbabayi,
  JHEP {\bf 1309}, 045 (2013)
  [arXiv:1305.6939 [hep-th]];
  R.~Barbieri,
  Phys.\ Scripta T {\bf 158}, 014006 (2013)
  [arXiv:1309.3473 [hep-ph]];
  A.~de Gouvea, D.~Hernandez and T.~M.~P.~Tait,
  arXiv:1402.2658 [hep-ph].

\bibitem{Farina:2013mla} 
  M.~Farina, D.~Pappadopulo and A.~Strumia,
  JHEP {\bf 1308}, 022 (2013)
  [arXiv:1303.7244 [hep-ph]].

\bibitem{Giudice:2013yca} 
  G.~F.~Giudice,
  arXiv:1307.7879 [hep-ph].

\bibitem{Foot:2013hna} 
  R.~Foot, A.~Kobakhidze, K.~L.~McDonald and R.~R.~Volkas,
  arXiv:1310.0223 [hep-ph].

\bibitem{Bardeen:1995kv}
  W.~A.~Bardeen,
  FERMILAB-CONF-95-391-T.

\bibitem{Wetterich:1983bi} 
  C.~Wetterich,
  Phys.\ Lett.\ B {\bf 140}, 215 (1984); 
  DESY-87-154.


\bibitem{Grisaru:1979wc}
  M.~T.~Grisaru, W.~Siegel and M.~Rocek,
  Nucl.\ Phys.\ B {\bf 159} 429 (1979).


\bibitem{Veltman:1980mj} 
  M.~J.~G.~Veltman,
  Acta Phys.\ Polon.\ B {\bf 12}, 437 (1981).

\bibitem{Jack:1989tv} 
  I.~Jack and D.~R.~T.~Jones,
  Phys.\ Lett.\ B {\bf 234}, 321 (1990);
  M.~S.~Al-sarhi, I.~Jack and D.~R.~T.~Jones,
  Z.\ Phys.\ C {\bf 55}, 283 (1992);
  M.~Chaichian, R.~Gonzalez Felipe and K.~Huitu,
  Phys.\ Lett.\ B {\bf 363}, 101 (1995)
  [hep-ph/9509223].


\bibitem{Hamada:2012bp} 
  Y.~Hamada, H.~Kawai and K.~-y.~Oda,
  Phys.\ Rev.\ D {\bf 87}, 053009 (2013)
  [arXiv:1210.2538 [hep-ph]];
  F.~Jegerlehner,
  arXiv:1304.7813 [hep-ph];
  F.~Jegerlehner,
  arXiv:1305.6652 [hep-ph];
  I.~Masina and M.~Quiros,
  arXiv:1308.1242 [hep-ph].

\bibitem{Kobakhidze:2013uoa} 
  A.~Kobakhidze,
  PoS ICHEP {\bf 2012}, 156 (2013)
  [arXiv:1303.5897 [hep-ph]].


\bibitem{Meissner:2007xv} 
  K.~A.~Meissner and H.~Nicolai,
  Phys.\ Lett.\ B {\bf 660}, 260 (2008)
  [arXiv:0710.2840 [hep-th]];
  H.~Aoki and S.~Iso,
  Phys.\ Rev.\ D {\bf 86}, 013001 (2012)
  [arXiv:1201.0857 [hep-ph]].

\bibitem{Meissner:2006zh}
  K.~A.~Meissner and H.~Nicolai,
  Phys.\ Lett.\ B {\bf 648} 312 (2007)
  [hep-th/0612165].



\bibitem{Foot:2007as}
  R.~Foot, A.~Kobakhidze and R.~R.~Volkas,
  Phys.\ Lett.\ B {\bf 655} 156 (2007)
  [arXiv:0704.1165 [hep-ph]];
  R.~Foot, A.~Kobakhidze, K.~L.~McDonald and R.~R.~Volkas,
  Phys.\ Rev.\ D {\bf 76} 075014 (2007)
  [arXiv:0706.1829 [hep-ph]].

\bibitem{Foot:2007iy}
  R.~Foot, A.~Kobakhidze, K.~L.~McDonald and R.~R.~Volkas,
  Phys.\ Rev.\ D {\bf 77} 035006 (2008) 
  [arXiv:0709.2750 [hep-ph]];
  R.~Foot, A.~Kobakhidze and R.~R.~Volkas,
  Phys.\ Rev.\ D {\bf 82} 035005 (2010)
  [arXiv:1006.0131 [hep-ph]].


\bibitem{Iso:2009ss} 
  S.~Iso, N.~Okada and Y.~Orikasa,
  Phys.\ Lett.\ B {\bf 676}, 81 (2009)
  [arXiv:0902.4050 [hep-ph]];
  S.~Iso, N.~Okada and Y.~Orikasa,
  Phys.\ Rev.\ D {\bf 80}, 115007 (2009)
  [arXiv:0909.0128 [hep-ph]];
  M.~Holthausen, M.~Lindner and M.~A.~Schmidt,
  Phys.\ Rev.\ D {\bf 82}, 055002 (2010)
  [arXiv:0911.0710 [hep-ph]];
  T.~Hur and P.~Ko,
  Phys.\ Rev.\ Lett.\  {\bf 106}, 141802 (2011)
  [arXiv:1103.2571 [hep-ph]];
  L.~Alexander-Nunneley and A.~Pilaftsis,
  JHEP {\bf 1009}, 021 (2010)
  [arXiv:1006.5916 [hep-ph]];
  R.~Foot, A.~Kobakhidze and R.~R.~Volkas,
  Phys.\ Rev.\ D {\bf 84}, 075010 (2011)
  [arXiv:1012.4848 [hep-ph]];
  R.~Foot and A.~Kobakhidze,
  arXiv:1112.0607 [hep-ph];
  K.~Ishiwata,
  Phys.\ Lett.\ B {\bf 710}, 134 (2012)
  [arXiv:1112.2696 [hep-ph]];
  J.~S.~Lee and A.~Pilaftsis,
  Phys.\ Rev.\ D {\bf 86}, 035004 (2012)
  [arXiv:1201.4891 [hep-ph]];
  N.~Okada and Y.~Orikasa,
  Phys.\ Rev.\ D {\bf 85}, 115006 (2012)
  [arXiv:1202.1405 [hep-ph]];
  S.~Iso and Y.~Orikasa,
  PTEP {\bf 2013}, 023B08 (2013)
  [arXiv:1210.2848 [hep-ph]];
  C.~Englert, J.~Jaeckel, V.~V.~Khoze and M.~Spannowsky,
  JHEP {\bf 1304}, 060 (2013)
  [arXiv:1301.4224 [hep-ph]].

\bibitem{Heikinheimo:2013fta}
  M.~Heikinheimo, A.~Racioppi, M.~Raidal, C.~Spethmann and K.~Tuominen,
  arXiv:1304.7006 [hep-ph];
  M.~Heikinheimo, A.~Racioppi, M.~Raidal, C.~Spethmann and K.~Tuominen,
  Nucl.\  Phys.\ B {\bf 876} 201 (2013)
  [arXiv:1305.4182 [hep-ph]];
  T.~Hambye and A.~Strumia,
  arXiv:1306.2329 [hep-ph];
  I.~Bars, P.~Steinhardt and N.~Turok,
  arXiv:1307.1848 [hep-th];
  M.~Heikinheimo, A.~Racioppi, M.~Raidal and C.~Spethmann,
  arXiv:1307.7146 [hep-ph];
  C.~D.~Carone and R.~Ramos,
  arXiv:1307.8428 [hep-ph];
  G.~Marques Tavares, M.~Schmaltz and W.~Skiba,
  Phys.\ Rev.\ D {\bf 89}, 015009 (2014)
  [arXiv:1308.0025 [hep-ph]];
  A.~Farzinnia, H.~-J.~He and J.~Ren,
  arXiv:1308.0295 [hep-ph];
  Y.~Kawamura,
  arXiv:1308.5069 [hep-ph];
  V.~V.~Khoze,
  arXiv:1308.6338 [hep-ph];
  E.~Gabrielli, M.~Heikinheimo, K.~Kannike, A.~Racioppi, M.~Raidal and C.~Spethmann,
  arXiv:1309.6632 [hep-ph];
  T.~G.~Steele, Z.~-W.~Wang, D.~Contreras and R.~B.~Mann,
  arXiv:1310.1960 [hep-ph];
  M.~Hashimoto, S.~Iso and Y.~Orikasa,
  arXiv:1310.4304 [hep-ph].

\bibitem{Holthausen:2013ota} 
  M.~Holthausen, J.~Kubo, K.~S.~Lim and M.~Lindner,
  arXiv:1310.4423 [hep-ph];
  S.~Abel and A.~Mariotti,
  arXiv:1312.5335 [hep-ph];
  C.~T.~Hill,
  arXiv:1401.4185 [hep-ph];
  J.~Guo and Z.~Kang,
  arXiv:1401.5609 [hep-ph];
  M.~Hashimoto, S.~Iso and Y.~Orikasa,
  arXiv:1401.5944 [hep-ph];
  B.~Radovcic and S.~Benic,
  arXiv:1401.8183 [hep-ph];
  A.~Salvio and A.~Strumia,
  arXiv:1403.4226 [hep-ph];
  J.~Kubo, K.~S.~Lim and M.~Lindner,
  arXiv:1403.4262 [hep-ph];
  V.~V.~Khoze, C.~McCabe and G.~Ro,
  arXiv:1403.4953 [hep-ph];
  G.~C.~Dorsch, S.~J.~Huber and J.~M.~No,
  arXiv:1403.5583 [hep-ph].

\bibitem{Shaposhnikov:2008xi} 
  M.~Shaposhnikov and D.~Zenhausern,
  Phys.\ Lett.\ B {\bf 671}, 162 (2009)
  [arXiv:0809.3406 [hep-th]];
  F.~Gretsch and A.~Monin,
  arXiv:1308.3863 [hep-th];
  C.~Tamarit,
  JHEP {\bf 1312}, 098 (2013)
  [arXiv:1309.0913 [hep-th]].


\bibitem{Coleman:1973jx}
  S.~R.~Coleman and E.~J.~Weinberg,
  Phys.\ Rev.\ D {\bf 7} 1888 (1973).

\bibitem{Vissani:1997ys} 
  F.~Vissani,
  Phys.\ Rev.\ D {\bf 57}, 7027 (1998)
  [hep-ph/9709409].

\bibitem{Volkas:1988cm} 
  R.~R.~Volkas, A.~J.~Davies and G.~C.~Joshi,
  Phys.\ Lett.\ B {\bf 215}, 133 (1988).

 

\end{thebibliography}
\end{document}